\documentclass[a4paper,11pt]{article}
\pdfoutput=1 

\usepackage{jcappub} 

\usepackage[utf8]{inputenc}
\usepackage{amsmath}
\usepackage{amssymb}
\usepackage{graphicx} 
\usepackage{bm}        
\usepackage{amssymb}   
\usepackage{mathtools}
\usepackage{xcolor}
\usepackage{hyperref}
\usepackage{orcidlink} 
\usepackage{xspace}
\usepackage{placeins}
\usepackage{mathrsfs}  
\usepackage{bigstrut}
\usepackage{soul}
\usepackage{ulem}

\begin{document}
\title{Constraining the Primordial Black Hole Abundance with Space-Based Detectors }

\author{Wencong Hong${}^{1,4}$\,\orcidlink{0009-0007-5010-3517}}\emailAdd{hongwencong@itp.ac.cn}
\author{Shi Pi${}^{1,2,3}$\,\orcidlink{0000-0002-8776-8906}}\emailAdd{shi.pi@itp.ac.cn} 
\author{Ao Wang${}^{1,4,5,6}$\,\orcidlink{0009-0002-6559-5212}}\emailAdd{wangao@itp.ac.cn}
\author{Zhenyu Zhang${}^{7}$\,\orcidlink{0000-0003-0869-4601}}\emailAdd{zhangzhenyu@nbu.edu.cn}

\affiliation{
    $^{1}$ Institute of Theoretical Physics, Chinese Academy of Sciences, Beijing 100190, China}
\affiliation{
    $^{2}$ Center for High Energy Physics, Peking University, Beijing 100871, China}
\affiliation{
    $^{3}$ Kavli Institute for the Physics and Mathematics of the Universe (WPI), The University of Tokyo, Kashiwa, Chiba 277-8583, Japan}
\affiliation{
    $^{4}$ School of Physical Sciences, University of Chinese Academy of Sciences, Beijing 100049, China}
\affiliation{
    $^{5}$ Institute for Theoretical Physics, Leibniz University Hannover, Appelstraße 2, 30167 Hannover, Germany}
\affiliation{
    $^{6}$ Department of Physics, Kyoto University, Kyoto 606-8502, Japan
}
\affiliation{
    $^{7}$ Institute of Fundamental Physics and Quantum Technology, and School of Physical Science and Technology,	Ningbo University, Ningbo, Zhejiang 315211, China}

\abstract{
Overdense regions can collapse into primordial black holes (PBHs) in the early universe, which are a compelling candidate for dark matter. Current constraints leave the asteroid-mass window the only possible one for PBH to account for all the dark matter, which can only be probed indirectly by the scalar-induced gravitational waves (GWs) sourced by the curvature perturbation which forms PBH. 
In this work, we explore the capabilities of future space-based gravitational wave detectors, including LISA, Taiji, and TianQin, to constrain such induced GWs as well as the PBH abundance. 
We systematically account for the width of the primordial curvature power spectrum, and find that the asteroid-mass window can be fully probed by all three space-based interferometers. If PBHs constitute the majority of dark matter, the induced GW leaves a strong signal in the mHz band 
with a signal-to-noise ratio of $10^3$--$10^4$. 
}

\maketitle

\section{Introduction}
The nature of dark matter (DM) remains one of the central 
problems in modern physics. Among the viable candidates, primordial black holes (PBHs), which can form through the direct gravitational collapse of rare, highly overdense regions in the early universe \cite{Zeldovich:1967lct,Hawking:1971ei,Carr:1974nx,Meszaros:1974tb,Carr:1975qj,Khlopov:1985jw}, have attracted growing attention due to their natural emergence without requiring new particles beyond the Standard Model. Unlike 
astrophysical black holes originated from remnants of stars, PBHs can span a wide mass range \cite{Green:2020jor}. 
Various experimental constraints have been reported in different mass ranges, 
including microlensing events \cite{MACHO:2000qbb,Tisserand:2006zx,Niikura:2017zjd,Mroz:2024mse,Mroz:2024wag}, gravitational wave (GW) mergers \cite{LVK:2022ydq,LIGOScientific:2025vwc}, and anomalous supernovae \cite{Smirnov:2022zip}. While none of these observations can allow PBH to fully account for dark matter, there are no robust constraint in the asteroid-mass window of $10^{-16}$--$10^{-10} M_\odot$. The femtolensing constraints \cite{Barnacka:2012bm,Katz:2018zrn,Jung:2019fcs} and dynamical constraints \cite{Capela:2012jz,Capela:2013yf,Defillon:2014wla,Graham:2015apa,Montero-Camacho:2019jte,Dai:2024guo} are either model-dependent or reply on future experiments (see \textit{e.g.} \cite{Carr:2020gox} for a review). Therefore, the asteroid-mass windows remains the only mass range where PBHs have not yet been observationally excluded as the dominant dark matter component, primarily due to their tiny size \cite{Niikura:2017zjd,Sugiyama:2019dgt}. 

A non-negligible population of PBHs typically requires sizeable primordial curvature perturbation, which unavoidably sources a stochastic gravitational wave background (SGWB) at quadratic order, known as scalar-induced gravitational waves (SIGWs) \cite{Tomita:1967wkp,Matarrese:1992rp,Matarrese:1993zf,Matarrese:1997ay,Ananda:2006af,Baumann:2007zm}. A rough estimate suggests that the required curvature perturbation amplitude is $\mathcal{P}_\zeta \sim 0.01$, leading to a current GW energy density of about $\Omega_{\mathrm{GW},0}h^2\sim10^{-6}\mathcal{P}_\zeta^2\sim10^{-10}$. 
A wide range of existing and upcoming gravitational wave experiments are capable of probing such a stochastic background, including pulsar timing arrays (PTAs) such as NANOGrav \cite{NANOGrav:2023gor,NANOGrav:2023hde}, EPTA \cite{EPTA:2023sfo,EPTA:2023fyk,EPTA:2023xxk}, PPTA \cite{Reardon:2023gzh,Zic:2023gta,Reardon:2023zen}, and CPTA \cite{Xu:2023wog}; space-based laser interferometers such as LISA \cite{LISA:2017pwj, Babak:2021mhe}, Taiji \cite{Luo:2019zal,Wu:2026tba}, and TianQin \cite{Luo:2025sos,Luo:2025ewp}; and next-generation space-based interferometers BBO \cite{Crowder:2005nr,Corbin:2005ny} 
and DECIGO \cite{Kawamura:2006up,Kawamura:2011zz}. With these experiments, we can probe PBHs across different mass ranges. The horizon size determines the peak frequency of SIGWs at the time of PBH formation, which also sets the PBH mass \cite{Saito:2008jc}
\begin{equation}\label{eq:f-M}
    f_\mathrm{IGW}\approx0.03 \mathrm{Hz}\left(\frac{M_\mathrm{PBH}}{10^{20}~\mathrm{g}}\right)^{-1/2}.
\end{equation}
This correspondence allows detectors at different frequencies to probe different PBH mass regions, for example, planet- to stellar-mass PBHs with PTAs \cite{Saito:2008jc,Byrnes:2018txb,Cai:2019elf,Chen:2019xse,Domenech:2020ers,Inomata:2023zup,Dandoy:2023jot,Domenech:2024rks,Franciolini:2023pbf,Iovino:2024tyg}. Notably, SIGWs associated with asteroid-mass PBHs fall within the sensitivity bands of LISA/Taiji/TianQin \cite{Saito:2008jc,Cai:2018dig,Unal:2018yaa,Bartolo:2018evs,Bartolo:2018rku,Byrnes:2018txb,Ren:2023yec,LISACosmologyWorkingGroup:2023njw,Luo:2025ewp,Iovino:2025xkq}. These missions thus offer a unique opportunity to probe the asteroid-mass PBH population, providing a direct test of this scenario in which PBHs make up a significant fraction of dark matter.
A simple comparison of $\Omega_\mathrm{GW}\sim10^{-10}$ with the noise curve gives a detectable frequency range of space-based interferometers $10^{-4}$--$10^{-1}$ Hz, which can be transferred to the PBH mass range of $10^{-13}$--$10^{-10} M_\odot$ by \eqref{eq:f-M}. This simple estimate leaves an undetectable gap in the asteroid-mass window $10^{-16}$--$10^{-13} M_\odot$. However, once we consider the spectral shape, especially the infrared tail of the SIGW, the detectable range can fully cover the asteroid-mass window. A detailed study of the detectability is necessary, especially when the spectral shape, the different noise curves of these interferometers, as well as constraints from the other experiments are considered. This is the main task of this paper.

This paper is organized as follows. In Section \ref{sec:PBH}, we review our method for estimating the PBH abundance, followed by a derivation of the SIGW energy spectrum from curvature perturbations parameterized by the primordial power spectrum in Section \ref{sec:SIGW}. In Section \ref{sec:detector}, we compare the design specifications and noise characteristics of three space-based interferometers, including the impact of confusion foregrounds. In Section \ref{sec:constraint}, we present the resulting constraints on PBHs in the asteroid-mass window and compare them with existing observational bounds. We conclude in Section \ref{sec:conclusion} by summarizing the implications of our findings for PBH dark matter.

\section{PBH abundance}
\label{sec:PBH}
The PBH abundance depends on the number of overdense regions and the probability of undergoing gravitational collapse. 
The former is determined by the statistical properties of primordial curvature perturbations, while the latter—owing to the highly nonlinear nature of the collapse—typically requires threshold values calibrated through numerical simulations, which are sensitive to the profiles of the peaks. In this section, we briefly introduce our method for calculating the abundance of primordial black holes, combining the Press–Schechter approach \cite{Young:2024jsu} with peak profiles in the broad-spectrum case \cite{Bardeen:1985tr,Pi:2024ert}.

\subsection{Peak Profile}

The requirement that PBHs do not overclose the universe implies that only rare, high-amplitude peaks in the curvature perturbations can undergo gravitational collapse, which are approximately spherically symmetric \cite{Bardeen:1985tr}. In this case, on superhorizon scales, the metric on the comoving slice can be expressed as 
\begin{equation}
\mathrm{d}s^2 = -\mathrm{d}t^2 + a^2(t) \mathrm{e}^{2\zeta(r)} \left( \mathrm{d}r^2 + r^2 \mathrm{d}\Omega^2 \right),
\end{equation}
where ${\rm d}\Omega^2 = {\rm d}\theta^2 + \sin^2\theta\,{\rm d}\phi^2$, and $\zeta$ is the comoving curvature perturbation. Given a power spectrum, one can derive the typical profile of high-density peaks \cite{Bardeen:1985tr,Yoo:2020dkz,Kitajima:2021fpq,Pi:2024ert}. However, as the universe evolves, perturbations on scales smaller than the Jeans scale $r_{\rm w}$ are suppressed by radiation pressure, which means the power spectrum will be different from the primordial one. Therefore we need a filtered power spectrum, which reflects the smoothing on different scales. Given the limited understanding of the nonlinear dynamics governing the evolution of high peaks, we introduce a Gaussian window function to model the smoothing effect on the curvature power spectrum  \cite{Ando:2018qdb,Young:2019osy,Young:2024jsu}:
\begin{equation}\label{eq:ps_wf}
\mathcal{P}_\zeta(k; r_{\rm w}) = \mathcal{P}_\zeta(k)\, \widetilde{W}^2(k; r_{\rm w}),\quad \widetilde{W}(k; r_{\rm w}) = \exp\left( -\frac{(kr_{\rm w})^2}{2} \right).
\end{equation}
At a given time, in the high-peak limit \cite{Pi:2024ert}, the most probable peak profile takes the form
\begin{equation}
\zeta(r; r_{\rm w}) = \mu\, \psi(r; r_{\rm w}),
\label{eq:HPL_prof}
\end{equation}
where $\mu$ denotes the peak amplitude and $\psi(r; r_{\rm w})$ is given by
\begin{equation}
    \psi(r;r_{\rm w}) = \frac{1}{\sigma^2(r_{\rm w})}\int k^2\mathcal{P}_\zeta(k;r_{\rm w})j_0(kr){\rm d\,ln}k,\quad \sigma^2(r_{\rm w})=\int\frac{{\rm d}k}{k}k^2\mathcal{P}_\zeta(k;r_{\rm w}),
\end{equation}
with $j_0(z) = \sin z/z$ the zeroth-order spherical Bessel function. Here, the probability of finding a peak with a given amplitude $\mu$ follows a specific statistical distribution, which will be discussed in detail in Section~\ref{sec:abund}.

\subsection{Threshold and PBH Mass}
The compaction function, first proposed in Ref.~\cite{Shibata:1999zs}, serves as a useful indicator for determining whether an overdense region will collapse \cite{Shibata:1999zs,Harada:2015yda}.  It is defined as the ratio of the Misner–Sharp mass excess to the areal radius:
\begin{equation}
\mathscr{C}(t,r) \equiv \frac{2\, G\, \delta M(t,r)}{a e^{\zeta(r)}r},
\label{eq:compaction difine}
\end{equation}
where $\delta M = M_\mathrm{MS}(r) - M_\text{B}$ represents the mass excess relative to the flat background enclosed within the same areal radius $a e^{\zeta(r)}r$. The compaction function can be further simplified as \cite{Harada:2015yda,Musco:2018rwt,Young:2019yug,Kawasaki:2019mbl}
\begin{align}
    \mathscr{C}(r)= \mathscr{C}_\ell-\frac{3}{8}\mathscr{C}_\ell^2,
    \label{eq:compaction defined by zeta}
\end{align}
where $\mathscr{C}_\ell(r)= -(4/3)r\,\partial_r\zeta$ is the linearized compaction function. For simplicity, we focus on the type-I fluctuations with $d (a\,r\,e^{\zeta})/d r>0$ or $\mathscr{C}_\ell < 4/3$.\footnote{For discussions on type-II fluctuation and related PBHs, see \textit{e.g.} \cite{Uehara:2024yyp,Shimada:2024eec,Uehara:2025idq,Escriva:2025ftp,Escriva:2025eqc,Escriva:2025rja}.} A PBH forms if the innermost local maximum of the compaction function, $\mathscr{C}_m \equiv \mathscr{C}(r_m)$ 
exceeds a threshold that depends on the peak profile. Numerical simulations show that this dependence can be captured by a functional form in terms of the shape parameter $q$ \cite{Escriva:2019phb,Musco:2020jjb,Germani:2023ojx}:
\begin{equation}
\mathscr{C}_{\rm th}(q) = \frac{4}{15}e^{-\frac{1}{q}} \frac{q^{1-\frac{5}{2q}}}{\Gamma\left(\frac{5}{2q}\right)-\Gamma\left(\frac{5}{2q},\frac{1}{q}\right)},
\end{equation}
where
\begin{equation}\label{eq:q}
q\equiv \left.\frac{-r^2\, \partial_r^2 \mathscr{C}(r)}{4 \mathscr{C}(r)\left(1 - \frac{3}{2} \mathscr{C}(r)\right)}\right|_{r=r_m}
\end{equation}
which roughly describes the width of the innermost maximum of $\mathscr{C}(r_m)$. In general, peaks which can collapse into PBHs are mainly concentrated near the so-called critical region, where the compaction function satisfies $\mathscr{C}_m - \mathscr{C}_{\rm th} \lesssim 10^{-2}$. In this regime, the PBH mass obeys a scaling law \cite{Choptuik:1992jv,Evans:1994pj,Koike:1995jm,Niemeyer:1997mt,Musco:2008hv}
\begin{equation}
    M_{\rm PBH}(\mathscr{C}_m)=K M_H\left(\mathscr{C}_m-\mathscr{C}_{\rm th}\right)^\gamma=K M_H\left(\mathscr{C}_{\ell,m}-\frac{3}{8}\mathscr{C}_{\ell,m}^2-\mathscr{C}_{\rm th}\right)^\gamma.
    \label{eq:PBH mass with horizon}
\end{equation}
$K$ and $\gamma$ are pure numbers determined by numerical relativity. For simplicity, we choose $K= 1$ and $\gamma=0.36$, following Ref.~\cite{Escriva:2021pmf}. The horizon mass can be written as \cite{Tada:2019amh,Kitajima:2021fpq}
\begin{equation}
    M_H(r_{\rm w})\approx 10^{20}~\mathrm{g} \left(\frac{g_*}{106.75}\right)^{-1/6}\left(\frac{r_{\rm w}}{6.41\times10^{-14}{\rm Mpc}}\right)^2\,,
\end{equation}
where $g_*$ is the relativistic degree of freedom at the moment of horizon reentry of the window scale $r_\mathrm{w}$.

\subsection{PBH abundance}\label{sec:abund}
\begin{figure}[h]
    \centering
    \includegraphics[width=0.5\textwidth]{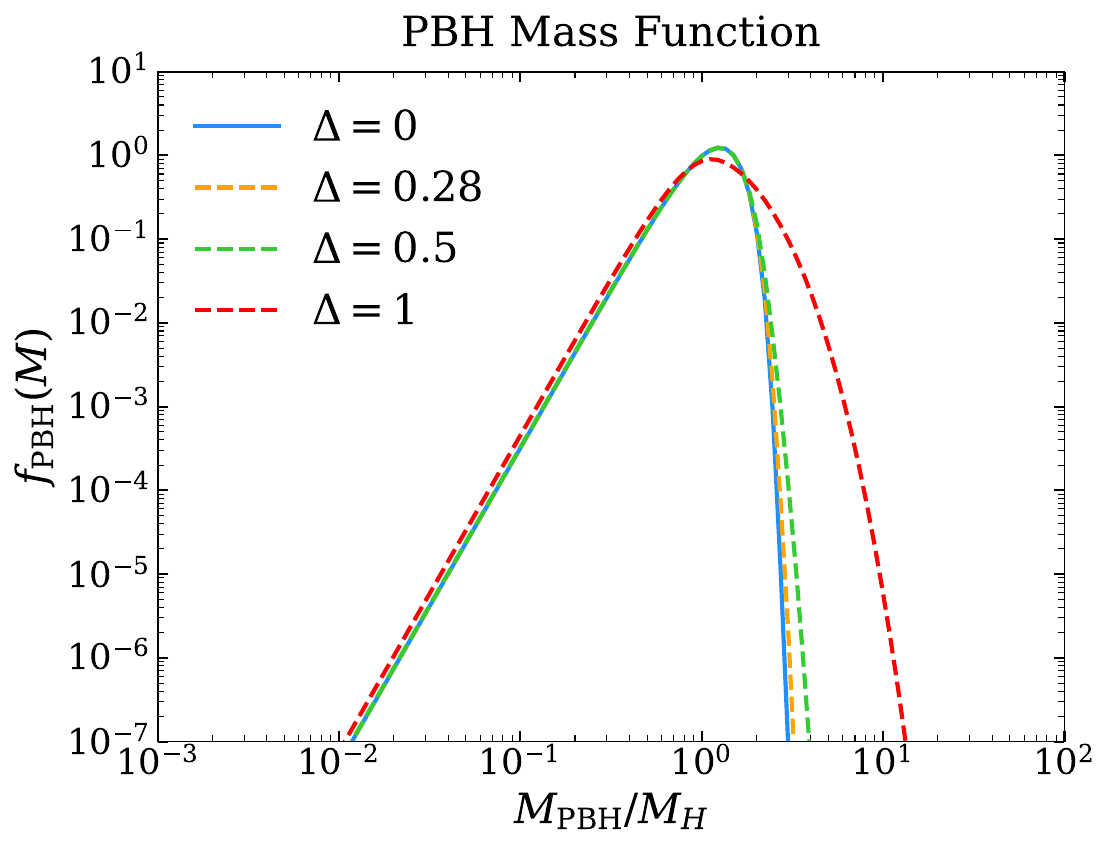}
    \caption{The PBH mass functions corresponding to power spectra with widths $\Delta = 0$ (blue), $0.28$ (orange), $0.5$ (green), and $1$ (red), all normalized to a total PBH abundance of $f_{\rm PBH} = 1$.}
    \label{fig:PBH mass function}
\end{figure}

Slow-roll inflation generically predicts an approximately scale-invariant spectrum of primordial curvature perturbations, with an amplitude of order $\zeta \sim 10^{-5}$ as measured in the CMB anisotropies by the Planck satellite \cite{Planck:2018jri}. The formation of PBHs requires a substantial enhancement of the curvature perturbations on smaller scales. The amplitude, spectral shape, and statistics of the enhanced curvature perturbations depend on the underlying mechanism. Leaving aside the model dependence, we adopt a widely used parameterization of the power spectrum in the log-normal form \cite{Pi:2020otn}:
\begin{equation}
    \mathcal{P}_\zeta(k) = \frac{A_\zeta}{\sqrt{2\pi}\Delta} \exp\left( -\frac{\ln^2(k/k_p)}{2\Delta^2} \right),
    \label{eq:ps}
\end{equation}
and assume that the curvature perturbations follow a Gaussian distribution.  Using Eq.~\eqref{eq:ps_wf}, we obtain the smoothed power spectrum of curvature perturbations at different times, characterized by the smoothing scale $r_{\rm w}$. Substituting this into Eq.~\eqref{eq:HPL_prof}, we derive the corresponding peak profile $\psi(r; r_{\rm w})$. For type-I fluctuations, the position of the compaction function peak, $r_m$, corresponds to the maximum of $r \partial_r \psi$ and thus depends on $r_{\rm w}$. By substituting Eq.~\eqref{eq:HPL_prof} into Eq.~\eqref{eq:compaction defined by zeta}, we obtain the compaction function $\mathscr{C}(r;r_{\rm w})$, from which we compute $q(r_{\rm w})$ via Eq.~\eqref{eq:q}. The distribution of the linearized compaction peak value $\mathscr{C}_{\ell,m}\equiv \mathscr{C}_{\ell}(r_m)$ then takes the form \cite{Kitajima:2021fpq,Gow:2022jfb,Pi:2024jwt}:
\begin{equation}
\mathbb{P}(\mathscr{C}_{\ell,m};r_{\rm w}) = \frac{1}{\sqrt{2\pi}\sigma_{\mathscr{C}_{\ell,m}}(r_{\rm w})} \exp\left(-\frac{\mathscr{C}_{\ell,m}^2}{2\sigma^2_{\mathscr{C}_{\ell,m}}(r_{\rm w})}\right),
\label{eq:probability of compact}
\end{equation}
where
\begin{equation}
    \sigma^2_{\mathscr{C}_{\ell,m}}(r_{\rm w})= \frac{16}{9}\int (kr_{\rm m})^2\mathcal{P}_\zeta(k;r_{\rm w})\left(\frac{{\rm d}j_0}{{\rm d}z}(kr_{\rm m})\right)^2{\rm d\, ln}k.
    \label{eq:rms of compact}
\end{equation}
Then, in the critical regime, the PBH abundance
at its formation is
\begin{align}\label{eq:beta}
    \beta(M_{\rm PBH};r_{\rm w})~{\rm d}\ln M_{\rm PBH} &=  \mathbb{P}(\mathscr{C}_{\ell,m};r_{\rm w}) \frac{M_{\rm PBH}}{M_H}\left(\frac{{\rm d\,ln}M_{\rm PBH}}{{\rm d}\mathscr{C}_{\ell,m}}\right)^{-1}\Theta(r_{\rm w}-\Xi r_m){\rm d\,ln}M_{\rm PBH}\notag\\
    &=K\frac{\left(\mathscr{C}_{\ell,m}-\frac{3}{8}\mathscr{C}_{\ell,m}^2-\mathscr{C}_{\rm th} (q(r_{\rm w}))\right)^{\gamma+1}}{\gamma\left(1-\frac{3}{4}\mathscr{C}_{\ell,m}\right)}\notag\\
    &\qquad\times\mathbb{P}(\mathscr{C}_{\ell,m};r_{\rm w})\Theta(r_{\rm w}-\Xi r_m)
    ~{\rm d\,ln}M_{\rm PBH},
\end{align}
where we have introduced an additional Heaviside step function $\Theta(r_{\rm w}-\Xi r_m)$ to avoid counting PBHs that are effectively formed at a different smoothing scale $r_{\rm w}$. The parameter $\Xi$ characterizes the relation between the size of the over-dense region and the horizon scale at the onset of collapse. In principle, $\Xi$ depends on the detailed peak profile and should be calibrated by numerical simulations. In this work, we adopt the same value $\Xi = 1/2.82$ as in Ref.~\cite{Pi:2024ert}. The PBH mass function normalized by the fraction of dark matter in the current universe is the sum of the contributions from PBHs formed at different times.
\begin{align}\nonumber
    f_{\rm PBH}(M_{\rm PBH})&\equiv \frac{1}{\Omega_{\rm dm}}\frac{{\rm d} \Omega_{\rm PBH}}{{\rm d}M_{\rm PBH}}\\
    & = 3.81\times10^8\left(\frac{h}{0.67}\right)^{-2}\left(\frac{g_{*,i}}{106.75}\right)^{-1/4}\notag\\
    &\qquad\times\int_0^\infty \frac{{\rm d}r_{\rm w}}{r_{\rm w}} 
    \left(\frac{M_H(r_{\rm w})}{M_\odot}\right)^{-1/2}\beta(M_{\rm PBH};r_{\rm w})~ \Theta\left(r_{\rm w} - \Xi r_m\right),
    \label{eq:fPBH with beta}
\end{align}
where $\left(M_H(r_{\rm w})/M_\odot\right)^{-1/2}$ represents the relative growth of PBH energy density with respect to radiation after formation. The dependence of the PBH mass function on the width of the power spectrum is illustrated in Fig.~\ref{fig:PBH mass function}.

\section{SIGW}
\label{sec:SIGW}
Large primordial curvature perturbations can induce second-order gravitational waves through mode coupling in the early universe (see Ref.~\cite{Domenech:2021ztg} for a review). It has recently been shown that tensor perturbations in the Newton gauge are directly related to the time delay measured by GW detectors \cite{Domenech:2025ccu}. Throughout this work, we therefore adopt this gauge and write the perturbed metric as
\begin{equation}
{\rm d}s^2 = a^2(\eta)\left[ -(1+2\Psi){\rm d}\eta^2 + \left( (1-2\Psi)\delta_{ij} + \frac{h_{ij}}{2} \right) {\rm d}x^i {\rm d}x^j \right]\,,
\label{eq:SIGW metric}
\end{equation}
where the tensor perturbation $h_{ij}$ is taken to be transverse and traceless (TT), i.e., $\delta^{ij} h_{ij} = 0$ and $\partial^i h_{ij} = 0$. Consider the polarization decomposition of tensor modes
\begin{equation}
h_{ij}(\eta,\mathbf{x}) = \sum_{\lambda=+,\times} \int \frac{{\rm d}^3k}{(2\pi)^3} e^{i\mathbf{k}\cdot\mathbf{x}} {\rm e}_{ij}^\lambda(\hat{\mathbf{k}}) h_\mathbf{k}^\lambda(\eta)\,,
\label{eq:polarization}
\end{equation}
where $+$ and $\times$ denote the two orthonormal polarization basis tensors that satisfy the TT gauge condition. Different polarization modes of the tensor perturbation satisfy the same evolution equation
\begin{equation}
{h_{\mathbf{k}}^{\lambda}}''(\eta) + 2\mathcal{H} {h_{\mathbf{k}}^{\lambda}}'(\eta) + k^2 h_{\mathbf{k}}^\lambda(\eta) = \mathcal{S}^{\lambda}(\eta,\mathbf{k})\,,
\label{eq:FT eom}
\end{equation}
where a prime indicates the derivative with respect to the conformal time $\eta$, and the TT part of the source term in Fourier space is given by
\begin{equation}
    \mathcal{S}^{\lambda}(\eta,\mathbf{k})=4 \int \frac{{\rm d}^3 q}{(2 \pi)^3} e_\lambda^{i j}(k) q_i q_j \Phi_{\mathbf{q}} \Phi_{|\mathbf{k}-\mathbf{q}|} f(\eta, q,|\mathbf{k}-\mathbf{q}|)\,.
\end{equation}
Here, the kernel $f$ related to the evolution of scalar perturbation can be found in Eq.~(3.20) in Ref.~\cite{Domenech:2021ztg}.
The equation of motion \eqref{eq:FT eom} can be solved by the Green function method, 
\begin{equation}
    h_\mathbf{k}^\lambda(\eta) = \frac{1}{a(\eta)} \int^\eta {\rm d}\tilde{\eta} \,G_k(\eta,\tilde{\eta}) a(\tilde{\eta}) \hat{\mathcal{S}}^{\lambda}(\mathbf{k},\tilde{\eta})\,,
    \label{eq:solution of GF}
\end{equation}
where $G_k(\eta,\tilde{\eta})=\sin\left(k(\eta-\tilde{\eta})\right)\Theta(\eta-\tilde{\eta})/k$ is the retarded Green function in the radiation dominated era. The resulting time-averaged power spectrum of SIGW is 
\begin{equation}
    \overline{\mathcal{P}_h}(\eta, k)=8 \int_0^{\infty} {\rm d}v \int_{|1-v|}^{1+v} {\rm d} u\left(\frac{4 v^2-\left(1-u^2+v^2\right)^2}{4 u v}\right)^2 \overline{I^2(\eta, k, v, u)} \mathcal{P}_{\Phi}(k u) \mathcal{P}_{\Phi}(k v)\,,
\end{equation}
where the kernel is defined as $I(\eta, k, u, v) \equiv \int^\eta d \tilde{\eta} ~G(\eta, \tilde{\eta}) f(\tilde{\eta}, k, u, v)$, which admits a closed-form analytical expression in the radiation-dominated era given by Eq.~(22) of Ref.~\cite{Kohri:2018awv}. (See also \cite{Espinosa:2018eve}.) In this era, the Bardeen potential $\Phi$ and the comoving curvature perturbation $\zeta$ are related by $\Phi = \frac{2}{3} \zeta$ for superhorizon modes $k\mathcal{H} \ll 1$. We can write down the energy spectrum of SIGWs both at the time of generation and at the present epoch
\begin{align}
    &\Omega_{{\rm GW}}(\eta_f,k) \equiv \frac1{\rho_{\rm cr}(\eta_f)}\frac{\mathrm{d} \rho_\mathrm{GW}}{\mathrm{d}\ln k} = \frac{1}{24}\left(\frac{k}{\mathcal{H}(\eta_f)}\right)^2 \overline{\mathcal{P}_h}(\eta_f,k),\\
    &\Omega_{\mathrm{GW},0}(k) h^2 = 1.62\times 10^{-5} \left(\frac{\Omega_{r,0}h^2}{4.18\times10^{-5}}\right)   \left(\frac{g_{*,S}(\eta_k)}{106.75}\right)^{-4/3} \left(\frac{g_{*}(\eta_k)}{106.75}\right)\Omega_{\mathrm{GW}}(\eta_f,k),
    \label{eq:Omega with PS}
\end{align} 
where $\eta_f\gg\eta_k=1/k$ is chosen such that the source term has sufficiently decayed. 
$g_{*,S}$ and $g_{*}$ denote the effective relativistic degrees of freedom for entropy and energy density, respectively, and $\rho_{\rm cr}(\eta_f)$ is the critical energy density of the universe at that time. In practice, for efficient and accurate computation of the SIGW spectrum, we utilize the \href{https://github.com/Lukas-T-W/SIGWfast/releases}{\textsc{SIGWfast}} package \cite{Witkowski:2022mtg}, which enables rapid scanning of the entire parameter space of the curvature perturbation power spectrum.

\section{Detection}
\label{sec:detector}
In this section, we review the detectability of SGWB signals in the space-based laser interferometers, including the Laser Interferometer Space Antenna (LISA), the Taiji Program in Space, and the TianQin Project. These detectors, through their unique designs and technological approaches, provide an opportunity to probe GW signals around the millihertz frequency band.

\begin{itemize}
    \item 
The Laser Interferometer Space Antenna (LISA) is a European Space Agency (ESA)–led space mission, developed in collaboration with NASA and member states, to detect low-frequency gravitational waves in the band from $10^{-4} {\rm Hz}$ to $10^{-1} {\rm Hz}$ \cite{LISA:2017pwj}. LISA will consist of three identical spacecraft flying in a triangular constellation with arm lengths of 2.5 million km in a heliocentric orbit trailing the Earth, enabling precision laser interferometry to measure perturbations of spacetime. The LISA Pathfinder technology demonstrator, launched on 3rd December 2015, successfully validated key technologies for space-based gravitational wave detection—achieving free-fall performance and interferometric precision that met and exceeded the mission’s requirements, thereby demonstrating the feasibility of the full observatory concept \cite{Armano:2018kix}. Following this success, LISA was formally adopted as ESA’s third large-class mission (L3) under the Cosmic Vision program in January 2024, advancing into detailed design and implementation; industrial contracts for spacecraft construction and key instrument development were signed in 2025, with a planned launch in the mid-2030s to commence science operations.

\item
The Taiji Program in Space is a Chinese space-based gravitational wave observatory initiative led by the Chinese Academy of Sciences to detect low-frequency gravitational waves in the band from approximately $10^{-4} {\rm Hz}$ to $10^{-1} {\rm Hz}$ using laser interferometry \cite{Luo:2019zal}. Taiji’s mission concept comprises three spacecraft forming an equilateral triangular constellation with arm lengths of 3 million km in a heliocentric orbit, enabling sensitivity to sources such as massive black hole mergers, extreme mass ratio inspirals, and stochastic backgrounds in the mid-to-low-frequency regime. Taiji Pathfinder (also known as Taiji-1), a technology demonstration satellite, was successfully launched in August 2019. In-orbit tests have verified key technologies, including laser interferometry at picometer precision, micro-thruster control at micro-Newton resolution, and gravitational reference sensing, providing critical validation for subsequent mission stages \cite{wu_chinas_2021}. Following this success, the Taiji program aims toward the full three-satellite constellation targeted for launch in the early 2030s to commence science operations.

\item
The TianQin Project is a Chinese space-based gravitational wave observatory proposed to detect low-frequency gravitational waves in the band from \(10^{-3}\,\mathrm{Hz}\) to \(1\,\mathrm{Hz}\)
\cite{TianQin:2015yph,TianQin:2020hid}. 
TianQin’s baseline design envisions three drag-free satellites in geocentric orbits with radii of \(\sqrt{3}\times10^5\,\mathrm{km}\) forming an equilateral triangle whose plane is oriented nearly perpendicular to the ecliptic, with the constellation’s sensitive axis pointing toward the ultracompact binary RX J0806.3+1527 as a reference source \cite{TianQin:2015yph,Luo:2016cjs,TianQin:2020hid}. To mature the required technologies, TianQin-1, a single satellite technology demonstrator launched in December 2019, has successfully tested drag-free control and inertial sensing beyond its mission requirements \cite{Luo:2020bls,Luo:2025sos}. The progress of the on-orbit validation of core subsystems demonstrates steady advancement toward the realization of TianQin’s scientific objectives.  
\end{itemize}

LISA, Taiji, and TianQin are all designed as triangular constellations consisting of three drag-free satellites forming an approximately equilateral triangle. This configuration allows the test masses to follow geodesic motion to high precision, thereby efficiently suppressing non-gravitational disturbances. As a consequence, the instrumental noise budget of these detectors is dominated by two fundamental contributions: the residual acceleration noise acting on the test masses and the displacement noise associated with the laser interferometric measurement \cite{Larson:1999we,Cornish:2002rt}. 
For a generic space-based interferometer labeled by the index $i$, the equivalent instrumental noise power spectral density can be written as \cite{Cornish:2002rt}
\begin{equation}
S_{{\rm ins},i}(f)=\frac{1}{\Gamma_i(f)L_i^2}\left(S_{x,i} + 2\left[1+ \cos^2\left(\frac{f}{f_{i}}\right)\right]\frac{S_{a,i}}{(2\pi f)^4}\right)\,,
\label{eq:Noise spectrum of ins}
\end{equation}
where $L_i$ is the arm length, $S_{x,i}$ is the displacement noise per one-way laser link, and $S_{a,i}$ is the residual acceleration noise per test mass along the sensitive direction. The characteristic transfer frequency is defined as $f_{*,i}=c/(2\pi L_i)$, and $\Gamma_i(f)=3\left[1+0.6(f/f_{*,i})^2\right]^{-1}/10$ represents the fitted overlap reduction function associated with the finite arm length and detector geometry. The specific design parameters adopted for LISA \cite{LISA:2017pwj}, Taiji \cite{Luo:2019zal}, and TianQin \cite{Luo:2025ewp} are summarized in Tab.~\ref{tab:detector param}.

\begin{table}[ht]
\centering
\caption{The Purpose Parameter}
\begin{tabular}{c c l l}
\hline
Project & Arm length & Displacement noise & Acceleration noise \\
& L(m) & $ S_x^{1/2}\rm (pm/Hz^{1/2})$ &$ S_a^{1/2}\rm (fm \cdot s^{-2} /Hz^{1/2})$\\
\hline
\multicolumn{4}{c}{} \\[-\normalbaselineskip]
LISA & $2.5\times 10^9$ & $ 15\cdot \sqrt{1+\left(\frac{2\,\rm mHz}{f}\right)^4} $ &
$3 \cdot \sqrt{1+\left(\frac{0.4\,\rm mHz}{f}\right)^2}\cdot \sqrt{1+\left(\frac{f}{8\,\rm mHz}\right)^4}$ \vspace{5pt}\\
Taiji & $3\times 10^9$ &$ 8\cdot \sqrt{1+\left(\frac{2\,\rm mHz}{f}\right)^4} $ & $3 \cdot \sqrt{1+\left(\frac{0.4\,\rm mHZ}{f}\right)^2}\cdot \sqrt{1+\left(\frac{f}{8\,\rm mHz}\right)^4}$ \vspace{5pt}\\
TianQin & $\sqrt 3\times 10^8$ & 1 & $1\cdot \sqrt{1+\frac{0.1\,\rm mHZ}{f}}$ \vspace{5pt}\\
\hline
\end{tabular}
\label{tab:detector param}
\end{table}

\begin{figure}[ht]
    \centering
    \includegraphics[width=0.8\textwidth]{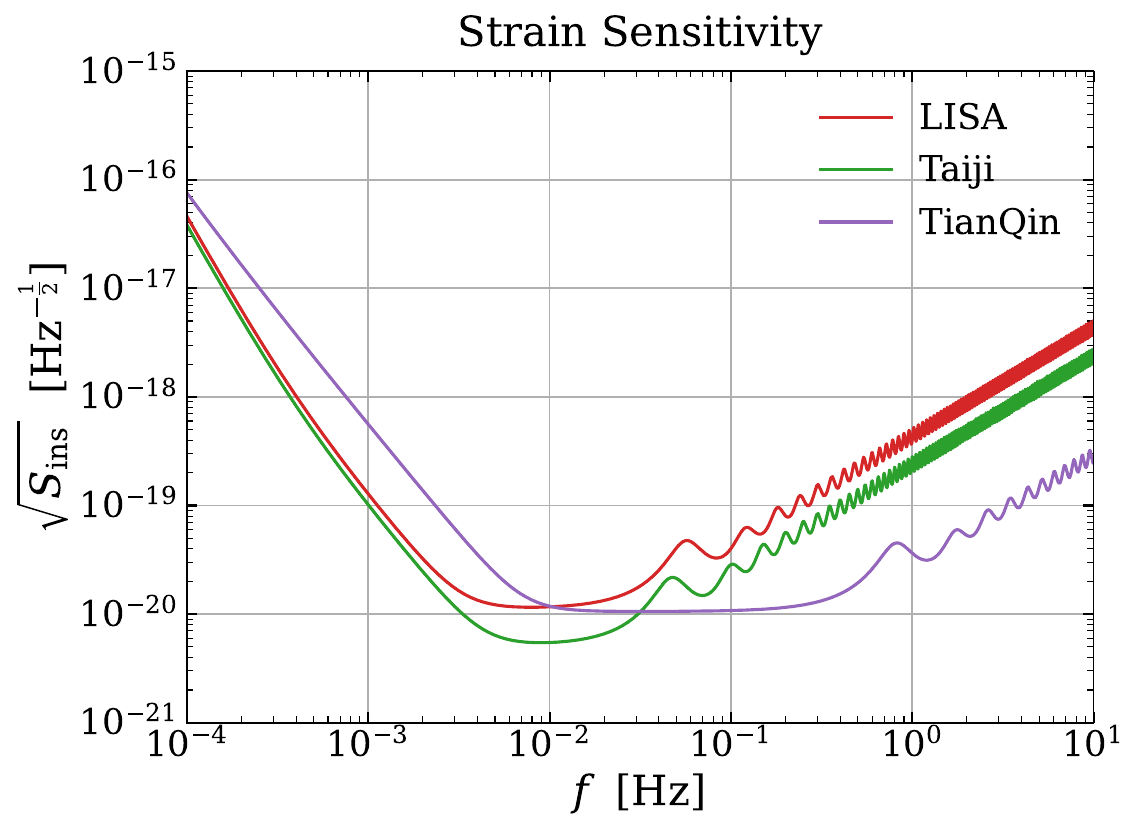}
    \caption{The strain sensitivity curves of LISA (red), Taiji (green), and TianQin (purple) for isotropic SGWB searches. The sensitivities are evaluated from the instrumental noise models \eqref{eq:Noise spectrum of ins} summarized in Tab.~\ref{tab:detector param}. Detectors with longer arm lengths achieve optimal sensitivity at lower frequencies. The low-frequency rise is dominated by residual acceleration noise, whereas the high-frequency degradation is caused by the reduced detector response.}
    \label{fig:detectors sensitivity}
\end{figure}

Using these parameters, we compute the strain sensitivity curves for the three detectors, which are shown in Fig.~\ref{fig:detectors sensitivity}. The overall minimum of each sensitivity curve, corresponding to the most sensitive frequency band, is primarily determined by the detector arm length. Detectors with longer arms probe lower frequencies more efficiently, as the characteristic transfer frequency $f_*$ scales inversely with $L$. This explains why LISA and Taiji achieve their optimal sensitivity at lower frequencies compared to TianQin. More specifically, in the low-frequency regime the acceleration noise contribution is amplified by the factor $(2\pi f)^{-4}$, which leads to a rapid degradation of the strain sensitivity as the frequency decreases. This steep low-frequency behavior represents a generic feature of space-based interferometers and sets a fundamental limitation on their sensitivity to gravitational waves on long time scales. In practice, the low-frequency performance differs slightly among LISA, Taiji, and TianQin due to their distinct acceleration noise models. To account for potential unknown disturbances at extremely low frequencies, the acceleration noise is commonly modeled with an additional low-frequency \textit{reddening} factor, such as $\sqrt{1+({0.1\,\rm mHz}/f)}$. As discussed in Ref.~\cite{ESA:2011NGO}, this precautionary treatment is motivated by the fact that the low-frequency behavior of the acceleration noise is extremely difficult to assess during ground-based testing and verification, and therefore represents a conservative design choice.

At high frequencies, the displacement noise becomes approximately frequency independent, while the detector response encoded in the overlap reduction function $\Gamma_i(f)$ decreases with increasing frequency. Consequently, the sensitivity curve rises again at high frequencies. Superimposed on this overall trend is a wiggly structure originating from the finite arm length of the detector, which leads to a frequency and direction dependent response to gravitational waves. When the gravitational-wave wavelength becomes comparable to or shorter than the arm length, the detector response is partially suppressed, producing the characteristic wavy features observed in the high-frequency sensitivity curves \cite{Cornish:2001qi}.

Meanwhile, the sensitivity of gravitational wave detectors depends not only on instrumental noise but also on astrophysical foreground \cite{Karnesis:2021tsh}. Within the Milky Way, tens of millions of compact binary systems, collectively known as galactic binaries, contribute to a significant foreground noise \cite{Lamberts:2019nyk}. These galactic binaries, which include white dwarf binaries and neutron star binaries, emit stochastic gravitational wave radiation. Characterized by orbital periods shorter than a few hours, these systems radiate GWs primarily in the frequency band of $0.1-10 \rm mHz$ \cite{Nelemans:2001hp,Nelemans:2000es}. This frequency range substantially overlaps with the sensitive bands of space-based detectors. Therefore, the gravitational wave background generated by white dwarf binaries is particularly noteworthy, as it acts as a prominent foreground noise that limits our ability to detect the stochastic gravitational wave background from the early universe. After accounting for individually resolvable binary events, the residual confusion foreground can be effectively modeled as a polynomial function in logarithmic frequency space
\begin{equation}
    \log_{10} S_{{\rm con},i}(f)=\sum_{k=0}^5 a_{k,i} \left[{\log}_{10}\left(\frac{f}{\rm 1\,mHz}\right)\right]^k\,.
\end{equation}
where the fitting coefficients $a_k$ are listed in Tab.~\ref{tab:foreground param} \cite{Wu:2023bwd}. This approach allows for a more accurate characterization of the foreground noise, thereby enhancing our capacity to isolate and study the underlying gravitational wave signals of interest.  Note that this empirical parametrization is only valid within the frequency range of $0.1$--$6 \,\rm mHz$.
\begin{table}[ht]
\centering
\caption{Fitting coefficients of the confusion foreground.}
\begin{tabular}{c r r r r r r}
\hline
Detector & $a_0$ & $a_1$ & $a_2$ & $a_3$ & $a_4$ &  $a_5$ \\
\hline
LISA & $-37.187$ & $-3.432$ & $-2.753$ & $-5.044$ &$-7.123$ & $-4.120$ \\
Taiji & $ -37.186$ &$-3.485$ &$-3.273$ &$-5.970$ & $-7.926$ & $-4.785$ \\
TianQin & $-37.262$ &$-3.465$ &$-2.790$ &$-2.128$ & $1.701$ & $2.734$ \\
\hline
\end{tabular}
\label{tab:foreground param}
\end{table}

With both the instrumental noise $S_{\rm ins}(f)$ and the astrophysical confusion foreground $S_{\rm con}(f)$ specified above, we now describe the strategy adopted to assess the detectability of the stochastic SIGW background. A SGWB is intrinsically random and does not admit a deterministic strain waveform. Instead, it is characterized statistically through its power spectrum. The signal measured by a space-based interferometer therefore consists of the superposition of several stochastic components, including the SIGW background of interest, the astrophysical confusion foreground, and the instrumental noise. In the triangular interferometer configuration employed by LISA, Taiji, and TianQin, the time-delay interferometry channels $A$ and $E$ are sensitive to gravitational waves, while the $T$ channel is approximately insensitive and can be used to monitor instrumental noise \cite{Smith:2019wny}. This allows for an effective subtraction of instrumental noise, but does not remove astrophysical foregrounds, which are themselves genuine gravitational-wave signals. Consequently, the confusion foreground cannot be treated on the same footing as instrumental noise in a standard signal-to-noise ratio (SNR) analysis.

We therefore adopt a conservative two-step strategy. The first step aims to quantify the intrinsic statistical sensitivity of each detector by considering instrumental noise as the only noise contribution, while treating all gravitational-wave backgrounds, including the astrophysical foreground, as signal components at the detector output level. In this step, the SNR is defined as
\begin{equation}
\varrho_i =
\left[
T \int {\rm d}f
\left(
\frac{\Omega_{\rm GW,0}(f) h^2}{\Omega_{{\rm ins},i}(f) h^2}
\right)^2
\right]^{1/2},
\label{eq:SNR of GW}
\end{equation}
with the instrumental noise spectrum
\begin{equation}
\Omega_{{\rm ins},i} h^2
= \frac{4\pi^2}{3H_{100}^2} f^3 S_{{\rm ins},i}(f)\,,
\end{equation}
where $\Omega_{\rm GW,0}(f)$ is the present-day energy density spectrum of the induced gravitational waves under consideration, $T$ denotes the total observation time, $i$ marks the different experiments, and $H_{100}=100\,\mathrm{km\cdot s^{-1}\cdot Mpc^{-1}}$. Throughout this work, we adopt $T = 3\,\mathrm{yr}$ and require a threshold SNR $\varrho_i = 3$ for the detection of SGWB. This procedure leads to the power-law-integrated (PLI) sensitivity curves \cite{Thrane:2013oya}, which visualize the response of the detector to a SGWB with a power-law spectrum.

As the second step, an additional physical requirement is imposed to ensure the identifiability of the induced signal in the presence of astrophysical foregrounds. Since the confusion foreground constitutes a genuine gravitational-wave background, the SIGW spectrum must exceed the foreground contribution in the relevant frequency range in order not to be overwhelmed in the measured data. Operationally, this requirement effectively raises the minimal detectable signal level in frequency bands dominated by the confusion foreground.

This criterion is motivated by Ref.~\cite{Caprini:2019pxz,Flauger:2020qyi,LISACosmologyWorkingGroup:2025vdz} on the reconstruction and interpretation of SGWB. In particular, even in analysis employing Bayesian or model-agnostic approaches, such as the power spectrum reconstruction method proposed in Ref.~\cite{LISACosmologyWorkingGroup:2025vdz}, where the mock data of SIGW signals can be successfully identified and reconstructed if they are not subdominant to the astrophysical foreground in the relevant frequency bands. In these situations, the reconstructed signal exhibits characteristic features of SIGWs, enabling a meaningful discrimination from other background components.

Within this framework, detectability is determined jointly by the statistical sensitivity set by instrumental noise and the astrophysical confusion foreground. Although simplified, this treatment yields a conservative estimate of detectability, which should be regarded as a lower bound in the presence of astrophysical confusion foregrounds.

\begin{figure}[ht]
    \centering
    \includegraphics[width=1\textwidth]{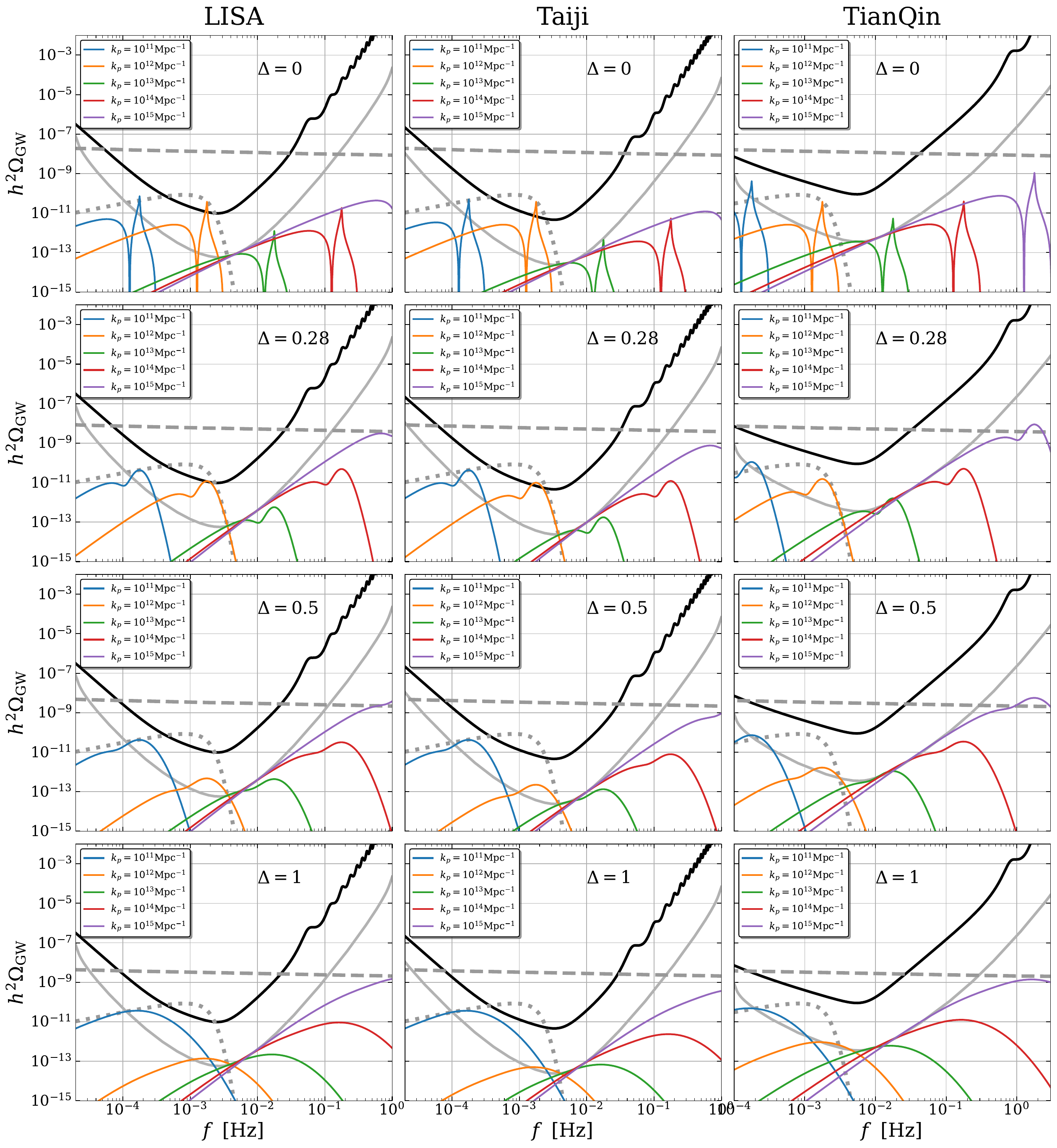}
    \caption{Noise spectrum for various space-based detectors, including the instrument noise (solid black) and the confusion foreground (dotted gray). The dashed gray line shows an envelope curve of $f_{\rm PBH} = 1$ obtained by shifting $\mathcal{A}_\zeta(k_p)$. Colored curves indicate the marginally detectable induced gravitational wave signals with SNR $\varrho_i=3$ for 3 years at different peak wavenumbers: $k_p = 10^{11}$ (blue), $10^{12}$ (orange), $10^{13}$ (green), $10^{14}$ (red), and $10^{15}$ (purple) $\mathrm{Mpc}^{-1}$.  From top to bottom, the rows correspond to power spectrum widths $\Delta = 0$, 0.28, 0.5, and 1, respectively. From left to right, the columns show the results for TianQin, LISA, and Taiji. The corresponding power-law-integrated sensitivity curves for SNR $\varrho_i = 3$ and observation time $T = 3\,\mathrm{yr}$ are shown in gray solid curves for comparison. }
    \label{fig:marginal SIGW and detectors}
\end{figure}

\section{Constraint on PBH Abundance}
\label{sec:constraint}
In this section, we discuss the constraints that various space-based detectors can place on the PBH abundance across different mass ranges. As shown in the previous section, these detectors can constrain the SGWB induced by the curvature perturbation at different frequencies, thereby limiting the amplitude of the curvature perturbation at different wavenumbers. 
\footnote{Primordial black hole binaries also emit GWs \cite{Sasaki:2016jop,Bird:2016dcv,Raidal:2018bbj,Raidal:2017mfl}, which form a SGWB by their superposition \cite{Ajith:2009bn,Ajith:2007kx,Mandic:2016lcn,Wang:2016ana,Bavera:2021wmw}. The peak of such a SGWB from asteroid-mass PBH binaries is higher than GHz, which is difficult to be probed and is not discussed in this paper.}
Consequently, these constraints translate into bounds on the existence of PBHs with masses approximately corresponding to the horizon mass at the time when these modes re-enter the horizon.

\begin{figure}[!t]
    \centering
    \includegraphics[width=1\textwidth]{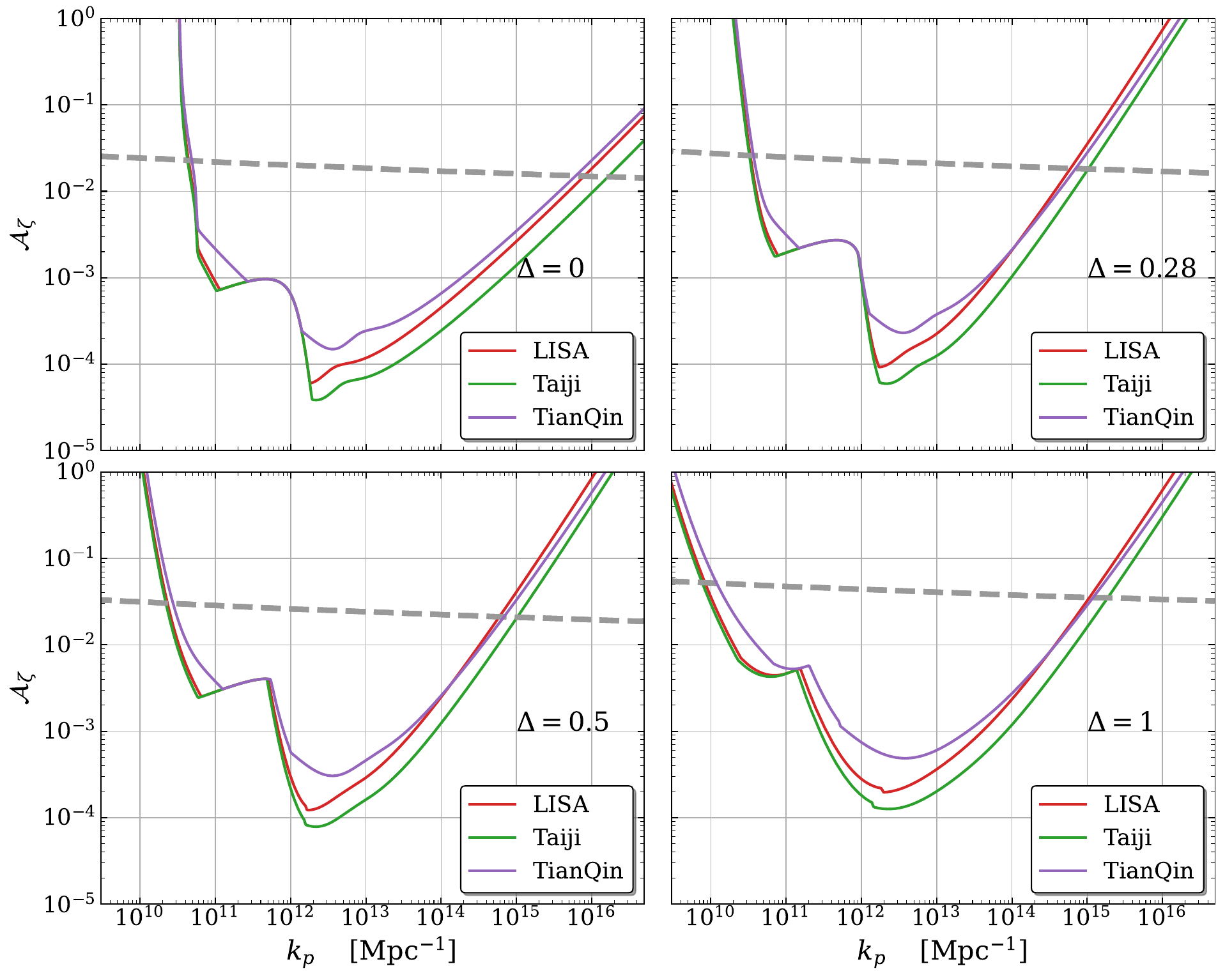}
    \caption{The constraints on the detectable region in amplitude $\mathcal{A}_\zeta$ and peak wavenumber $k_p$ for power spectra with $\Delta = 0$ (upper left), $\Delta = 0.28$ (upper right), $\Delta = 0.5$ (lower left), $\Delta = 1$ (lower right). The solid curves represent the sensitivity of LISA (red), Taiji (green) and TianQin (purple), with a total duration of $3\,\mathrm{yr}$ and SNR $\varrho = 3$. The gray dashed line represents $f_{\rm PBH} = 1$.}
    \label{fig:detectors' constraint on Ak vs kp}
\end{figure}

\begin{figure}[ht]
    \centering
    \includegraphics[width=1\textwidth]{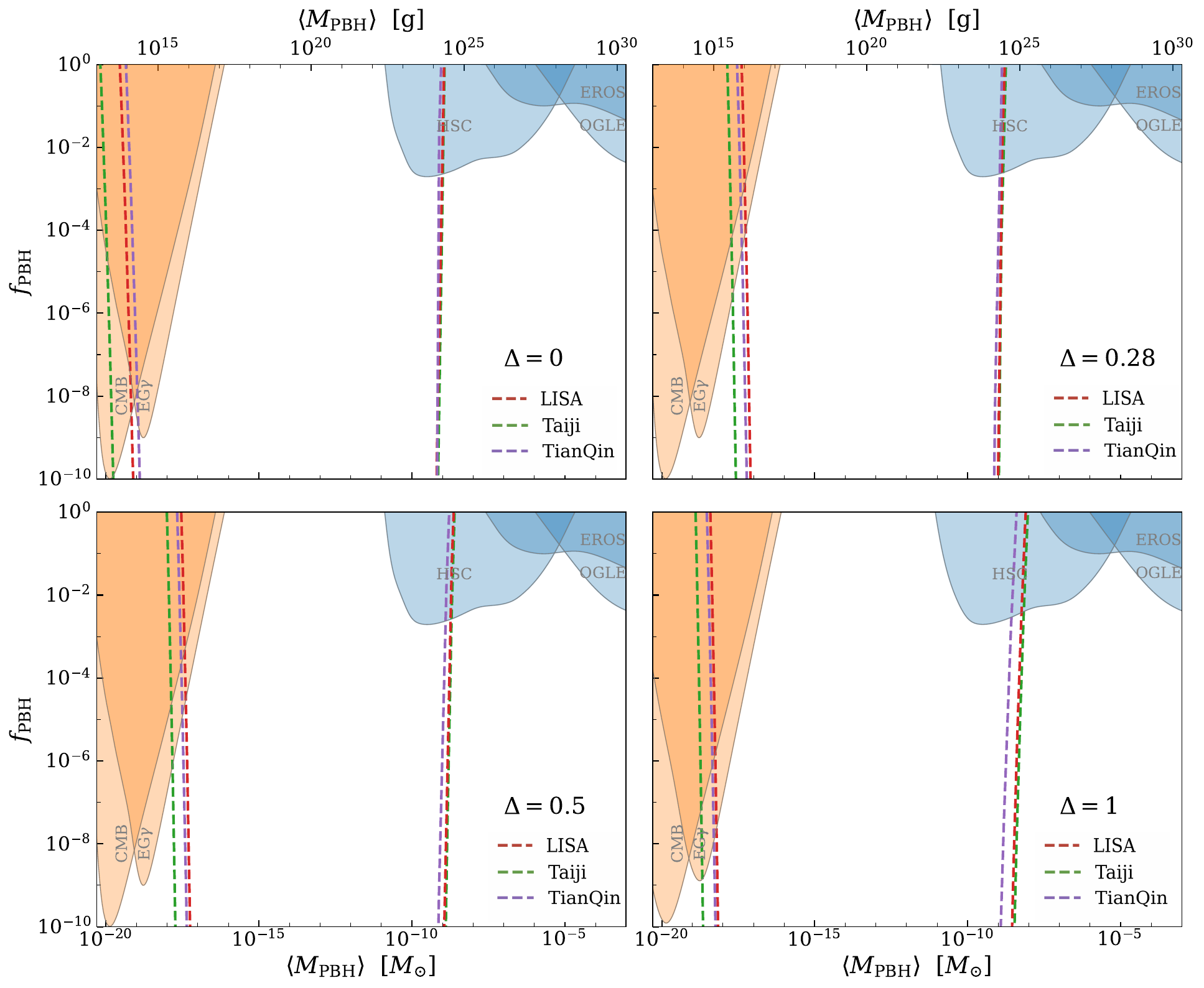}
    \caption{Constraints from different observations on the fraction of PBH in DM, $f_{\rm PBH}$, as a function of the PBH average mass $\langle M_{\rm PBH} \rangle$, for PBH mass function shown in Fig.~\ref{fig:PBH mass function} with monochromatic spectrum $\Delta = 0$ (upper left) and log-normal spectrum $\Delta = 0.28$ (upper right), $\Delta = 0.5$ (lower left) and $\Delta = 1$ (lower right). The orange region on the left is excluded by constraints from evaporation effects on CMB anisotropies \cite{Acharya:2020jbv} and extragalactic $\gamma$-rays \cite{Carr:2020gox}. The right blue region is excluded by microlensing (blue shades) of Subaru HSC \cite{Niikura:2017zjd}, EROS \cite{EROS-2:2006ryy}, and OGLE \cite{Mroz:2024mse}. The region between colored dashed lines is the detectable parameter space of LISA (red), Taiji (green), and TianQin (purple), with a total duration of $3 \mathrm{yr}$ and SNR $\varrho = 3$.}
    \label{fig:detectors' constraint on fPBH vs MPBH}
\end{figure}

The range of detectable wavenumbers of the curvature perturbations depends not only on the peak amplitude of the SIGW $\Omega_{\rm GW}(k_p)\propto \mathcal{A}_\zeta^2$, but also on its ultraviolet (UV) and infrared (IR) tails. If the curvature perturbation spectrum is monochromatic, the IR tail of the SIGW spectrum is $k^2$ \cite{Kohri:2018awv}. As the width of the curvature perturbation spectrum $\Delta$ increases, the infrared spectrum has a broken power-law, from $k^3$ ($k\lesssim k_p\Delta$) in the far-IR to $k^2$ ($k\gtrsim k_p\Delta$) in the near-IR \cite{Pi:2020otn}, which approaches a universal $k^3$ scaling as $\Delta\to\mathcal{O}(1)$ \cite{Cai:2019cdl}. 
Therefore, a spectrum peaked at frequencies higher than millihertz band is still possible to be probed, thanks to such IR tails. The detectable $A_\zeta(k)$ is naturally stretched to the higher wavenumbers, which extends the detectable PBH mass range, covering the aforementioned ``mass gap'' of $10^{-16}$ -- $10^{-13}~M_\odot$. On the other hand, according to $\Omega_\mathrm{GW}\propto\mathcal{P}_\zeta^2$ and  \eqref{eq:ps}, the UV tail of the SIGW spectrum is also an exponential decay, with a smaller width $\Delta/\sqrt2$ \cite{Pi:2020otn}. 
As a result, the lower bound of the detectable wavenumber in $A_\zeta(k)$ is nearly a sharp cutoff, and thereby makes the upper bound of the detectable PBH mass $\sim10^{-11}\,M_\odot$ robust, which is consistent with the rough estimation from the simple frequency-mass relation \eqref{eq:f-M}.
These features can be clearly seen in Fig.~\ref{fig:detectors' constraint on Ak vs kp} and Fig.~\ref{fig:detectors' constraint on fPBH vs MPBH}.

Next we convert the constraints on $A_\zeta(k)$ to those on the PBH abundance $f_\mathrm{PBH}(M_\mathrm{PBH})$. Using the formalism reviewed in Sec.~\ref{sec:PBH}, the wavenumber can be mapped to the mass of PBH, while the SIGW amplitude determines the PBH abundance.  The extended PBH mass function, illustrated in Fig.~\ref{fig:PBH mass function}, can be normalized as 
\begin{equation}
    f_{\rm PBH}(M_{\rm PBH}) = f_{\rm PBH} F_{\Delta}(M_{\rm PBH}/\langle M_{\rm PBH} \rangle),
\end{equation}
where their abundance and mass can be characterized as \cite{Andres-Carcasona:2024wqk}
\begin{align}
    f_{\rm PBH}&= \int {\rm d}\ln M_{\rm PBH}~ f_{\rm PBH}(M_{\rm PBH}),\\
    \langle M_{\rm PBH} \rangle &= f_{\rm PBH}\left(\int \frac{{\rm d}\ln M_{\rm PBH}}{M_{\rm PBH}}f_{\rm PBH}(M_{\rm PBH})\right)^{-1}.
    \label{eq:mean PBH mass}
\end{align}
Here the normalized mass function $F_\Delta$ can be regarded as a function of the width $\Delta$ of the power spectrum. This is because, when realizing different values of $\langle M_{\rm PBH}\rangle$ and $f_{\rm PBH}$, the amplitude $\mathcal{A}_\zeta$ remains nearly unchanged over a wide region of parameter space.

\begin{figure}[ht]
    \centering
    \includegraphics[width=1\textwidth]{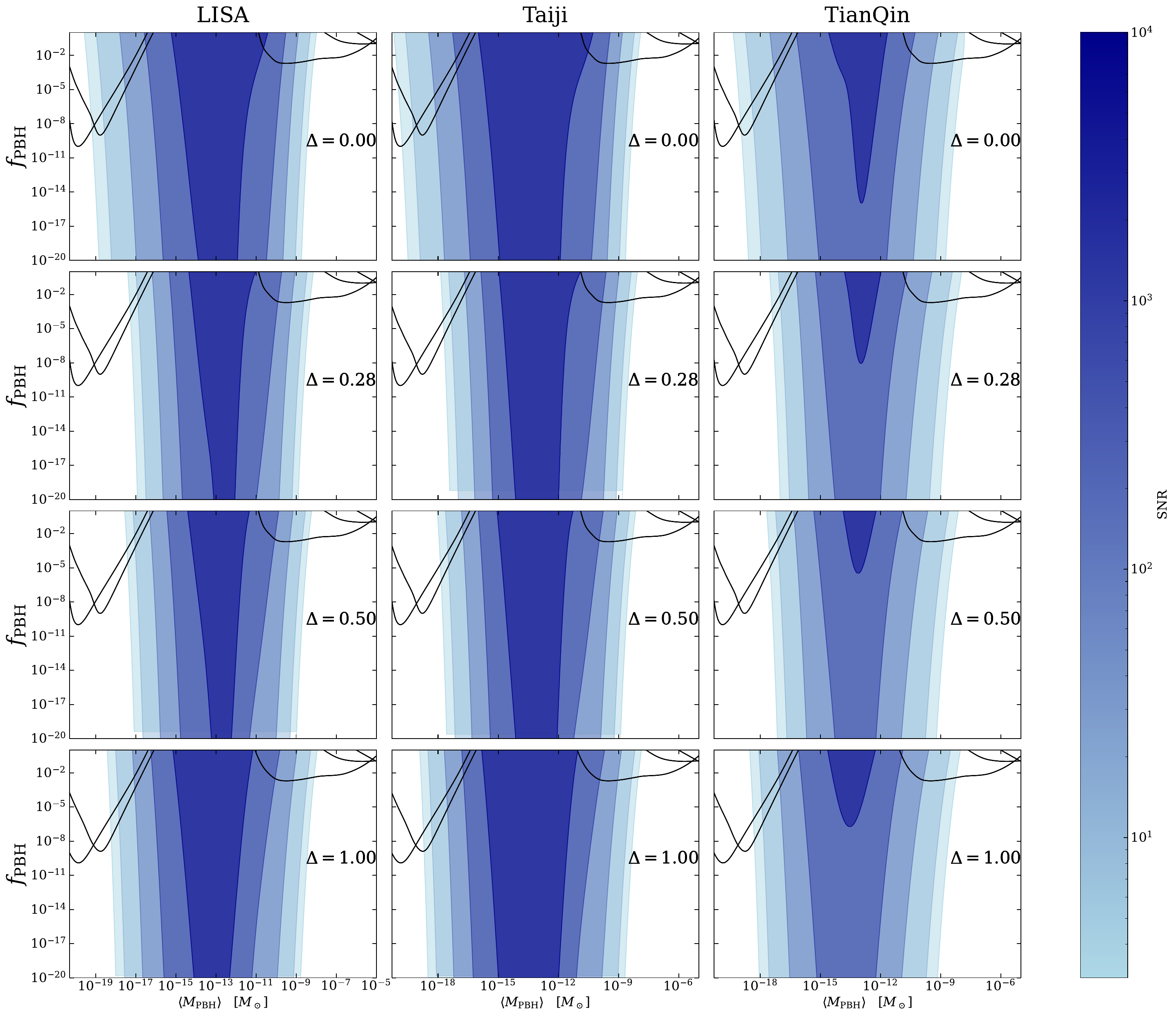}
    \caption{The contour plots of the SNR for induced GWs detected by the LISA (left), Taiji (middle), and TianQin (right), corresponding to different widths of the curvature perturbation spectra $\Delta$ (0, 0.28, 0.5 and 1). The black solid lines in the upper left and right corners correspond to the same constraints  in Fig.~\ref{fig:detectors' constraint on fPBH vs MPBH}. The contours are shown in shades of blue, with increasing color intensity indicating larger signal-to-noise ratios. From outer to inner, the contours correspond to $\mathrm{SNR}=3,\,10,\,10^2,\,10^3,$ and $10^4$, respectively.}
    \label{fig:SNR contour on fPBH vs MPBH}
\end{figure}

In Fig.~\ref{fig:detectors' constraint on fPBH vs MPBH}, we present the resulting constraints for power spectra with different widths, together with comparisons to various limits from evaporation and microlensing experiments. The latter constraints, applied to PBHs with an extended mass function centered at $\langle M_{\rm PBH} \rangle$, are derived from the bounds on PBHs with a monochromatic mass function $f_{\rm mo}(M_{\rm PBH})$ \cite{Carr:2017jsz}, by imposing the condition
\begin{equation}
\int {\rm d}\ln M_{\rm PBH}\,\frac{f_{\rm PBH}F(M_{\rm PBH}/\langle M_{\rm PBH} \rangle)}{f_{\rm mo}(M_{\rm PBH})} \leqslant 1\,,
\end{equation}
which ensures that the total contribution from PBHs across all masses does not exceed the observational limit. We note that the constraints from microlensing and evaporation remain nearly unchanged, since they depend only on the shape of the normalized mass function $F_\Delta$, which varies only mildly, as shown in Fig.~\ref{fig:PBH mass function}. As mentioned before, for these power spectra, all three interferometers LISA, Taiji, and TianQin can cover the entire PBH-dark-matter mass window sandwiched between microlensing and evaporation constraints. Furthermore, we notice that the lower bounds from space-based detectors, given by IR tails from higher-frequency peaks, depend sensitively on the width of the power spectrum.

\begin{figure}[ht]
    \centering
    \includegraphics[width=0.5\textwidth]{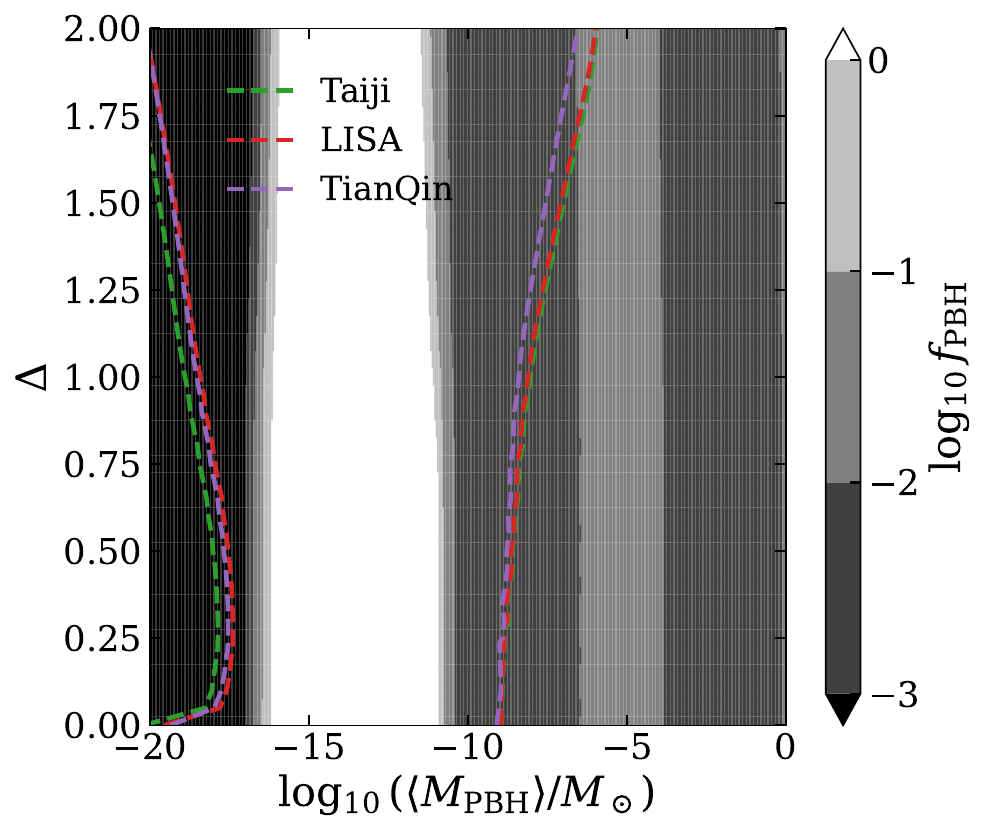}
    \caption{Combined constraints in the $\langle M_{\rm PBH}\rangle$–$\Delta$ plane for a log-normal curvature perturbation spectrum with width $\Delta$. The shaded background indicates the current upper limits on the PBH dark-matter fraction $f_{\rm PBH}$ as a function of PBH mass and spectral width $\Delta$, derived from reported evaporation and microlensing constraints for a monochromatic spectrum. Darker regions correspond to stronger bounds, with ${\log}_{10} f_{\rm PBH}<-3$ in the black region, while the white region remains weakly constrained by current observations ${\log}_{10} f_{\rm PBH}<0$. The colored dashed curves show the sensitivity reaches of space-based gravitational-wave detectors, assuming ${\rm SNR}=3$ and an observation time of $T_{\rm obs}=3\,\mathrm{yr}$. The regions enclosed by each pair of dashed curves represent the PBH mass ranges that can be probed under the assumption $f_{\rm PBH}=1$ for a given $\Delta$. Green, red, and purple curves correspond to Taiji, LISA, and TianQin, respectively. 
    }
    \label{fig:contour on Delta vs MPBH}
\end{figure}

To illustrate this dependence on the power-spectrum width, we draw a contour plot of $f_\mathrm{PBH}$ on the $\Delta$–$\langle M_{\rm PBH} \rangle$ plane in Fig.~\ref{fig:contour on Delta vs MPBH}, in which we combine current observational bounds using
\begin{equation}
    \sum_{j=1}^N\left(\int \mathrm{d} \ln M_{\rm PBH}\,\frac{f_{\rm PBH}F(M_{\rm PBH}/\langle M_{\rm PBH} \rangle)}{f_{{\rm mo},j}(M_{\rm PBH})}\right)^2 \leqslant 1,
\end{equation}
where $f_{{\rm mo},j}(M_{\rm PBH})$ denotes the $j$-th observational constraint. The resulting joint constraints are shown as a grayscale background, with darker regions indicating stronger exclusion. For clarity, we show only the $f_{\rm PBH} = 1$ contour from SIGW detection.
We notice that the width of the detectable PBH mass range is not a monotonic function of $\Delta$, and there is a minimum at $\Delta\approx0.28$. This is because, when $\Delta\ll1$, the IR scaling is $k^2$, which allows the GW peak to go to higher frequencies. On the other hand, $\Delta\approx\mathcal O(1)$, the width of the $\mathcal{P}_\zeta$ itself is roughly the width of $\Omega_\mathrm{GW}$, which makes the detectable range bigger. We have to emphasize that even for the minimal detectable window with $\Delta \approx 0.28$, it is still sufficient to cover the entire asteroid-mass window, guaranteeing that SIGW associated with PBHs will be detectable if dark matter is composed of asteroid-mass PBHs. 
Actually, as there is still parameter space beyond the asteroid-mass window, the SNR of the SIGW signal when PBHs are all the dark matter can be as high as $10^3$--$10^4$, as is clearly shown in  Fig.~\ref{fig:SNR contour on fPBH vs MPBH}.

This demonstrates that the capability of space-based interferometers to probe asteroid-mass PBH dark matter is robust against the choice of detection threshold, owing to the typically large signal-to-noise ratios expected for the SIGW signals, thereby enhancing the reliability of prospective detections against systematic uncertainties and more conservative detection requirements.

\section{Conclusion and Discussions}
\label{sec:conclusion}

Probing PBHs as a dark matter candidate is among the key scientific objectives of future space-based gravitational-wave interferometers such as LISA, Taiji, and TianQin. In this work, we have quantitatively assessed the capability of these detectors to constrain the PBH abundance over a broad mass range centered on the asteroid-mass window, which remains largely unconstrained by existing observations. Taking into account the dependence on the shape of the primordial curvature power spectrum, we find that all space-based interferometers considered here are expected to detect the associated induced gravitational-wave signal with signal-to-noise ratios of order $10^{3}$--$10^{4}$ across the entire mass window, assuming that PBHs constitute all of the dark matter, and largely independent of the specific form of the power spectrum.

The amplitude of SIGWs to be probed is normalized by the abundance of PBHs, which depend on an accurate calculation of their abundance. Two main approaches are widely used in the literature: the peak theory and the Press–Schechter formalism, of which the former systematically predicts a higher abundance \cite{Kitajima:2021fpq,Pi:2024ert}. 
In this work, we adopt the peak profile from peak theory and compute the abundance using the Press–Schechter method, following the prescription of \cite{Biagetti:2021eep,Kitajima:2021fpq,Gow:2022jfb,Pi:2024jwt},
as we prefer to avoid subtle issues like properly treating sub-Jeans scale suppression due to radiation pressure, its impact on the resulting PBH mass, the consistent summation of contributions from PBHs formed at different times, etc. Such uncertainties can affect both the overall PBH abundance and the dependence of the mass function on the shape of the primordial power spectrum. Nevertheless, we expect the qualitative features of our results to remain robust. Notably, numerical simulations suggest that the parameter $K$ in Eq.~\eqref{eq:PBH mass with horizon}, which relates the PBH mass to the horizon mass, depends on the pressure gradient determined by the peak profile, and often uses the horizon radius $r_m$ instead of the Jeans radius $r_{\rm w}$ \cite{Escriva:2021pmf,Musco:2023dak}. Compared to the fiducial choice $K = 1$ adopted in this work, a more accurate treatment may yield PBH masses up to an order of magnitude larger, potentially shifting the low-mass edge of the detectable region beyond the constraints of Hawking radiation. However, this gap is expected to be closed soon as more X-ray and gamma-ray data are considered from PBH evaporation \cite{DeRocco:2019fjq,Laha:2019ssq,Laha:2020ivk,Iguaz:2021irx,Mittal:2021egv}. For simplicity, we have also neglected the effects of accretion and mergers after PBH formation, which may further modify the mass distribution. More accurate quantitative predictions will require further theoretical and numerical developments.

It is worth emphasizing that the PBH abundance is highly sensitive to the statistical properties of overdense regions, which may be affected by both the non-Gaussianity of curvature perturbations and the statistical formalism used to evaluate the number density and the profiles of peaks. Non-Gaussian features naturally arise in many inflation models that can enhance the power spectrum, such as ultra-slow-roll inflation \cite{Namjoo:2012aa, Chen:2013eea, Cai:2018dkf, Biagetti:2018pjj, Passaglia:2018ixg, Pi:2022ysn, Artigas:2024ajh, Ballesteros:2024pbe,Cruces:2024pni,Wang:2024wxq,Cruces:2025typ,Caravano:2025diq, Escriva:2025ftp}, constant-roll inflation \cite{Atal:2019cdz,Atal:2019erb,Atal:2019erb,Atal:2019erb,Ragavendra:2023ret,Domenech:2023dxx,Wang:2024xdl,Inui:2024fgk,Shimada:2024eec}, the curvaton scenario \cite{Sasaki:2006kq,Pi:2021dft,Ferrante:2022mui,Franciolini:2023pbf,Hooper:2023nnl}, stochastic inflation \cite{Pattison:2021oen,Asadi:2023flu,Tomberg:2023kli,Raatikainen:2023bzk,Ballesteros:2024pwn,Jackson:2024aoo,Murata:2025onc,Briaud:2025ayt,Nassiri-Rad:2025dsa}, etc. For instance, if non-Gaussian curvature perturbation can be expanded perturbatively, at quadratic order, we have $\zeta=\zeta_{\rm G}+(3/5)f_{\mathrm{NL}}\zeta_{\rm G}^2$.
A positive $f_{\mathrm{NL}}$ enhances the PBH formation, which requires smaller curvature perturbation to generate the same amount of PBHs, thus predicts smaller SIGW signal. Up to quadratic level, such a suppression is mild, and PBH dark matter centered in the asteroid-mass window can still be probed by LISA, Taiji, and TianQin with a smaller signal-to-noise ratio
\cite{Cai:2018dig,Pi:2024lsu}. However, the low-mass edge of the window might be undetectable when $f_\mathrm{NL}$ is large \cite{Iovino:2025cdy}. For logarithmic non-Gaussianity of the form $\zeta = -(1/\gamma)\ln(1-\gamma\zeta_{\mathrm G})$ with $\gamma \lesssim 3.1$ (corresponding to a nominal $f_{\mathrm{NL}} \lesssim 2.6$), the suppression of the SIGW signal is still mild \cite{Abe:2022xur, Inui:2024fgk}. For larger values of $\gamma$, however, PBHs formed through the bubble channel become dominant \cite{Escriva:2023uko, Escriva:2025ftp}, thereby leading to a further suppression of the associated SIGW signal. A detailed investigation of this regime is therefore left for future work.

\vspace{1em}
\textbf{Note Added:} While this paper was in preparation, \cite{Iovino:2025cdy} appeared on arXiv, which studies how to probe PBH dark matter with LISA. The local non-Gaussianity of the curvature perturbation is also taken into account. Our result for LISA in the Gaussian case is in agreement with \cite{Iovino:2025cdy}, and our paper also includes the analysis of Taiji and TianQin.

\section*{Acknowledgements}
We thank Sachiko Kuroyanagi for early collaboration, and thank Jaume Garriga and Jia-ning Wang for helpful comments and discussions. We also thank Huai-Ke Guo and Jian-dong Zhang for revising our discussion on the recent progress of Taiji and TianQin, respectively. S.P. thanks Chenggang Shao and the Center for Gravitational Experiments of Huazhong University of Science and Technology for the hospitality during his visit, and thanks the APCTP workshop ``New Perspectives on Cosmology 2026'' (APCTP-2026-F02). This work is supported by the National Key Research and Development Program of China Grant No. 2021YFC2203004, and JSPS KAKENHI grant No. 24K00624. S.P. is also supported by the National Natural Science Foundation of China (NSFC) Grant Nos. 12475066 and 12447101. A.W. is also supported by the DFG under the Emmy-Noether program project number 49659236 and the UCAS Joint PhD Training Program. Z.Z. is also supported by NSFC Grant Nos. 12275004 and 12547127.

\bibliography{main}
\bibliographystyle{apsrev4-2}
\end{document}